\input phyzzx.tex
\tolerance=1000
\voffset=-0.0cm
\hoffset=0.7cm
\sequentialequations
\def\rl{\rightline}

\def\t1{{\tilde 1}}

\def\t{\theta}

\REF{\ABR}{E. R. C. Abraham and P. K. Townsend, Phys. Lett. {\bf B 291} (1992) 85; Phys. Lett. {\bf B 295} (1992) 225.}
\REF{\WIT}{E. Witten, Nucl. Phys. {\bf B 507} (1997) 658, [arXiv:hep-th/9706109].}
\REF{\GPT}{J. P. Gauntlett, R. Portugues, D. Tong and P. K. Townsend, Phys. Rev. {\bf D 63} (2001) 085002, [arXiv:hep-th/0008221].}
\REF{\ACH}{B. S. Acharya and C. Vafa, [arXiv:hep-th/0103011].}
\REF{\TONG}{D. Tong, [arXiv:hep-th/0509216].}
\REF{\AMI}{A. Hanany and D. Tong, JHEP {\bf 0404} (2004) 066, [arXiv:hep-th/0403158]; Comm. Math. Phys. {\bf 266} (2006) 647, [arXiv:hep-th/0507140].}
\REF{\NOS}{Y. Isozumi, M. Nitta, K. Ohashi and N. Sakai, Phys. Rev. Lett. {\bf 93} (2004) 161601, [arXiv :hep-th/0404198]; Phys. Rev. {\bf D70} (2004) 125014, [arXiv:hep-th/0405194];
Phys. Rev. {\bf D71} (2005) 065018, [arXiv:hep-th/0405129].}
\REF{\TON}{D. Tong, Phys. Rev. {\bf D66} (2002) 025013, [arXiv:hep-th/0202012].}
\REF{\YAN}{N. Sakai and Y. Yang, Comm. Math. Phys. {\bf 267} (2006) 783, [arXiv:hep-th/0505136].}
\REF{\LEE}{K. Lee and H. U. Yee, Phys. Rev. {\bf D72} (2005) 065023, [arXiv:hep-th/0506256].}
\REF{\ETO}{M. Eto, Y. Isozumi, M. Nitta and K. Ohashi, Nucl. Phys. {\bf B 52} (2006) 140, [arXiv:hep-th/0506257].}
\REF{\KAK}{K. Kakamoto and N. Sakai, Phys. Rev. {\bf D68} (2003) 065005,[arXiv:hep-th/0306077].}
\REF{\EIN}{M. Eto, Y. Iozumi, M. Nitta, K. Ohashi and N. Sakai, Phys. Rev. {\bf D 72} (2005) 085004, [arXiv:hep-th/0506135]; Phys. Lett. {\bf B632} (2006) 384, [arXiv:hep-th/0508241].}
\REF{\MET}{M. Eto et. al., Phys. Rev. {\bf D71} (2005) 125006, [arXiv:hep-th/0412024].}
\REF{\METO}{M. Eto et. al., Phys. Rev. {\bf D76} (2007) 105002, arXiv:0704.2218[hep-th].}
\REF{\SHI}{M. Shifman and A. Yung, Phys. Rev. {\bf D67} (2003) 125007, [arXiv:hep-th/0212293]; Phys. Rev. {\bf D70} (2004) 025013, [arXiv:hep-th/0312257].}
\REF{\DVA}{G. Dvali and A. Vilenkin, Phys. Rev. {\bf D67} (2003) 046002, [arXiv:hep-th/0209217].}
\REF{\SAT}{N. Sakai and D. Tong, JHEP {\bf 0503} (2005) 019, [arXiv:hep-th/0501207].}
\REF{\DAV}{D. Tong, JHEP {\bf 0602} (2006) 030, [arXiv:hep-th/0512192].}
\REF{\ASY}{R. Auzzi, M. Shifman and A. Yung, Phys. Rev. {\bf D 72} (2005) 025002, [arXiv;hep-th/0504148].} 
\REF{\LAST}{E. Halyo, arXiv:0906.3472[hep-th].}
\REF{\SING}{F. Cachazo, K. Intriligator and C. Vafa, Nucl. Phys. {\bf B603} (2001), [arXiv:hep-th/0103067]; F. Cachazo, S. Katz and C. Vafa, [arXiv:hep-th/0108120]}
\REF{\SUP}{E. Witten, Nucl. Phys. {\bf B507} (1997) 658, [arXiv:hep-th/9706109]; M. Aganagic and C. Vafa, [arXiv:hep-th/0012041].}
\REF{\SHA}{M. Aganagic, C. Beem and S. Kachru, Nucl. Phys. {\bf B796} (2008) 1, arXiv:0709.4277[hep-th].}
\REF{\DOU}{M. Douglas, JHEP {\bf 9707} (1997) 004, [arXiv:hep-th/9612126].}
\REF{\JAU}{D. E. Diaconescu, M. R. Douglas and J. Gomis, JHEP {\bf 9802} (1998) 013, [arXiv:hep-th/9712230].}
\REF{\SUS}{E. Halyo, arXiv:0906.2377[hep-th].} 
\REF{\COS}{E. Halyo, JHEP {\bf 0403} (2004) 047, [arXiv:hep-th/0312268].}
\REF{\CST}{E. Halyo, arXiv:0906.2587[hep-th].}


\singlespace
\rl{SU-ITP-09-48}
\pagenumber=0
\normalspace
\medskip
\bigskip
\titlestyle{\bf{Domain Walls on Singularities}}
\smallskip
\author{ Edi Halyo{\footnote*{e--mail address: halyo@stanford.edu}}}
\smallskip
\centerline {Department of Physics} 
\centerline{Stanford University} 
\centerline {Stanford, CA 94305}
\smallskip
\vskip 2 cm
\titlestyle{\bf ABSTRACT}

We describe domain walls that live on $A_2$ and $A_3$ singularities. The walls are BPS if the singularity is resolved and non--BPS if it is deformed and fibered.
We show that these domain walls may interpolate between vacua that support monopoles and/or vortices.

\singlespace
\vskip 0.5cm
\endpage
\normalspace

\centerline{\bf 1. Introduction}
\medskip

Theories with discrete vacua contain codimension one topological defects, i.e. domain walls, which interpolate between
two different vacua[\ABR]. Depending on the properties of these vacua, domain walls can shed light on the semi--classical and
nonperturbative physics in these theories. Domain walls have been investigated in great detail in (supersymmetric) field theories and those arising from intersecting brane configurations
[\WIT-\ASY]. 

On the other hand, domain walls may also live in theories obtained from wrapping branes on singular spaces such as $A_n$ singularities. 
In ref. [\LAST] it was shown that solitons of different dimensions, and in particular domain walls may live on such singularities. The world--volume theories of D5 branes wrapped on an 
$A_n$ singularity usually have isolated supersymmetric vacua which give rise to domain walls. In the brane configuration, these domain walls
correspond to D5 branes wrapped on a node of the singularity and stretched to a neighboring node along $C(x)$, the two dimensions transverse to $A_n$. The wrapping and stretching of the D5 brane
produce exactly the tension of the wall obtained in the world--volume field theory.

In this paper, we consider domain walls that live on the world--volumes of D5 branes which are wrapped on $A_2$ and $A_3$ singularities. If the singularities are
resolved but not deformed, we find that domain walls are BPS. If the singularity is deformed and fibered over $C(x)$, the ${\cal N}=2$ supersymmetry is broken and we obtain 
non--BPS domain walls. 
We find that, in certain cases, these walls interpolate between vacua with very different perturbative and nonperturbative properties; e.g. vacua supporting monopoles and/or vortices. In
such cases, domain walls may be connected by vortices which gives rise to configurations similar to those of D--branes connected by strings.
Singularities with a large number of nodes, i.e. $A_n$ with large $n$, 
lead to a large number of isolated vacua and therefore to a large number of domain walls with different properties in the world--volume theory.

Since our main aim in this paper is to show the existence of different types of domain walls that live on singular spaces, we do not investigate their quantum properties such as their 
interactions, 
world--volume theories, moduli spaces etc. This has been done in detail for domain walls that arise in intersecting brane models[\AMI,\MET,\METO]. We expect the quantum properties 
of our domain walls to be similar to but somewhat different from those in [\AMI,\MET,\METO] since 
our brane configurations either have only ${\cal N}=1$ supersymmetry or in the cases with ${\cal N}=2$ supersymmetry are not exactly dual to the intersecting brane models considered in 
those works. 

The paper is organized as follows. In section 2, we decribe BPS domain walls that live on a resolved $A_2$ singularity which is not deformed. In section 3, we describe non--BPS domain
walls that live on deformed $A_2$ and $A_3$ singularities which are fibered over $C(x)$. Section 4 includes a discussion of our results and our conclusions.

\bigskip
\centerline{\bf 2. BPS Domain Walls}
\medskip

In this section, we describe domain walls in ${\cal N}=2$ supersymmetric models that live on singularities. Due to the  ${\cal N}=2$ supersymmetry, these are 
BPS domain walls. The simplest model with such a domain wall is obtained from D5 branes wrapping an $A_2$ singularity given by
$$uv=(z-z_0)(z-z_0)(z-z_0) \eqno(1)$$
where $z_0$ may or may not vanish. The $A_2$ singularity has two nodes ($S^2$s) that overlap at the origin of the two dimensional (complex) transverse space $C(x)$.
We wrap $N_f$ D5 branes on the first node and $N_c$ D5 branes on the second one. On the 3+1 dimensional noncompact world--volume,
this leads to a $U(N_f) \times U(N_c)$ gauge theory with two chiral fields $\phi_1, \phi_2$ in the $(adj_f,1)$ and $(1,adj_c)$ representations respectively[\SING]. In
addition, there are two bifundamental chiral fields $Q_{12},Q_{21}$ in the $({\bar N_f},N_c)$ and $(N_f,{\bar N_c})$ representations respectively. These fields are coupled 
through the superpotential[\SUP,\SHA]
$$W=Tr[Q_{12}Q_{21}(\phi_2-\phi_1)] \eqno(2)$$
Clearly, for $Q_{12}=Q_{21}=0$, $\phi_1$ and $\phi_2$ are free; in fact this is the Coulomb branch of the theory. At a generic point in the Coulomb branch, we have (since we can always
take $\phi_2=0$ which simply defines the origin of the moduli space) 
$\phi_1=diag(a_1,a_2, \ldots, a_N)$ and $U(N_f)$ is broken to $U(N_f) \to U(1)^{N_f}$ with all Abelian couplings equal to $g_f$. We can decouple the gauge and singlet fields that live 
on the first node from the rest of matter by taking $g_f <<1$. The gauge coupling $g_f$ is given by
$${{4 \pi} \over g_{f}^2}={V_1\over {(2 \pi)^2 g_s \ell_s^2}} \eqno(3)$$
where $g_s$ and $\ell_s$ are the string coupling and length respectively. $V_1$ is the ``stringy volume''[\DOU] of the first node given by $V_1=(2 \pi)^4 \ell_s^4(B_1^2+r_1^2+\alpha_1^2)$ with
$$B_1=\int_{S^2} B^{NS}   \qquad  r_1^2=\int_{S^2} J \eqno(4)$$
i.e. $B_1$ is the NS-NS flux through the first node and $r_1^2$ is the volume of the $S_1^2$. We see that we can obtain $g_f <<1$ by taking $V_1 >>\ell_s^2$, e.g. by assuming 
a very large NS-NS flux through the first node, $S_1^2$. This decouples $\phi_1$ (with the frozen nonzero VEVs along the Coulomb branch) and turns  
$U(1)^{N_f}$ into a global symmetry. Thus, we are left with a $U(N_c)$ gauge group with $N_f$ flavors, $Q_{12i}$ and $Q_{21i}$ (in the $N_c$ and ${\bar N_c}$ representations respectively) 
and the adjoint $\phi_2$. The superpotential for these fields is given by
$$W=Tr[Q_{12}Q_{21}(\phi_2-diag(a_1,a_2, \ldots, a_{N_f}))] \eqno(5)$$
In addition, we can blow up the second node so that $S_2^2$ has a finite volume $r_2^2$. On the world--volume theory, this introduces an anomalous D--term $r_2^2=\xi$ for the $U(1)$[\JAU]
subgroup of $U(N_c)$ and its D--term becomes 
$$D=\sum_{i=1}^{N_f}(|Q_{12i}|^2-|Q_{21i}|^2+\xi I_{N_c}) \eqno(6)$$

In order to have supersymmetric vacua we need to assume $N_c \leq N_f$ since for $N_c>N_f$, $|Q_{21i}|^2$ which is an $N_f \times N_f$ matrix is not large enough to cancel the 
anomalous D--term which is an $N_c \times N_c$ matrix. The above theory has isolated vacua given by the different choices of $N_c$ VEVs which are the diagonal elements of $\phi_2$;
$\phi_2=diag(b_1,b_2, \dots, b_{N_c})$. A vacuum is obtained when one or more of the $N_c$ diagonal $b_j$s match one or more of the $N_f$ diagonal entries $a_i$; i.e. a vacuum is a collection
$C_k$ of $N_c$ entries out of $N_f$ possible values $a_i$. The number of these vacua is $N_f!/(N_f-N_c)!N_c!$. For a particular choice of $C_k$ with $b_j=a_i$ for a particular set of $i$, the 
corresponding $Q_{21i}$ are free and can obtain VEVs to cancel the anomalous D--term so that
$$\sum_j |Q_{21j}|^2=\xi  \delta_{ij}\qquad  \qquad Q_{12}=0  \qquad i=1, \ldots, N_f \eqno(7)$$
Due to the existence of isolated vacua, the theory contains domain walls which interpolate between any two different vacua given by two sets $C_k \not= C_{k^{\prime}}$. These are 
non--Abelian domain walls and their solutions, in general, are quite cumbersome to write explicitly. 

For simplicity let us consider the solution for the simplest case with $N_c=1$ which corresponds to Abelian domain walls. In this case, $\phi_2$ is a singlet with  
a VEV, $b$ that may be equal to any of the $N_f$ $a_i$s. Thus there are $N_f$ isolated vacua with $b=a_i$ for $i=1, \ldots, N_f$. Consider a domain wall interpolating between vacua
with VEVs $b=a_i$ and $b=a_j$. The solution for such a domain wall (in the limit $g_c^2 \xi>>a_i^2$) is given by
$$Q_{21i}={\sqrt{\xi} \over A} e^{-|a_i-a_j|z} \qquad Q_{21j}={\sqrt{\xi} \over A} e^{-|a_i-a_j|z} \eqno(8)$$
where $A^2=e^{-2|a_i-a_j|z}+e^{2|a_i-a_j|z}$ and
$$\phi_{2}={{a_i+a_j} \over 2}+{{a_i-a_j} \over 2}tanhz  \eqno(9)$$
In eqs. (8) and (9), for simplicity, we assumed the walls to be normal to the $z$ direction and located at $z=0$. Due to the ${\cal N}=2$ supersymmetry of the world--volume theory, these domain 
walls are BPS. Therefore, their solution and tension given by $T_w = \xi (|a_i-a_j|) $ do not receive any corrections.

In ref. [\LAST], it was shown that these domain walls are D5 branes wrapped on the blown up second node and stretched between the two (wrapped) D5 branes at $a_i$ and $a_j$ along the transverse 
plane $C(x)$. Since the D5 branes that form the domain walls are transverse to the three world wolume directions they form two dimensional surfaces as required.
The wrapping on the node contributes the factor $\xi$ in the tension whereas the factor $|a_i-a_j|$ arises from the stretching along $C(x)$. The details of this description 
can be found in ref. [\LAST] and will not be repeated here.

Consider the general case with $N_c$ wrapped D5 branes. In a generic vacuum, the gauge symmetry is broken to $U(N_c) \to U(1)^n \times U(k)$ where $n+k=N_c$ and $U(k)$ is spontaneously
broken by the bifundamental VEVs. Such a vacuum supports ($n$ types of) monopoles due to the unbroken Abelian symmetries and non--Abelian vortices due to the spontaneously broken $U(k)$.
A generic monopole has a mass of $m_m=4 \pi |a_i-a_j|/g^2$ and is a D3 brane wrapped on the second node that stretches from $a_i$ to $a_j$ on the complex plane $C(x)$[\LAST]. In addition, there
are $k$ types of vortices (since a non--Abelian vortex is obtained by embedding an Abelian vortex in the diagonal subgroup of $U(k)$) with tension $T=2 \pi \xi$. These vortices are
D3 branes wrapped on the second node that stretch along the (noncompact) world--volume[\LAST,\COS,\CST]. It has been shown that these vortices may end on the domain walls giving a 
configuration similar 
to that of D--branes with strings stretched between them. This similarity has been noticed in refs. [\SHI-\ASY]] where it was shown that the vortices are charged under the wall gauge group 
and the whole configuration is BPS. This raises the interesting possibility of a field theory that has a limit in which domain walls and vortices behave like to D--branes and strings. 


\bigskip
\centerline{\bf 3. Non--BPS Domain Walls}
\medskip

We can also obtain non--BPS domain walls in theories with ${\cal N}=1$ supersymmetry that live on a singularity. Even though these walls are not BPS they are expected to be stable 
since they interpolate between
two isolated vacua and are therefore topologically stable. In order to break supersymmetry down to ${\cal N}=1$ we need to wrap the D5 branes on a deformed $A_2$ fibered over $C(x)$. 
The simplest such singularity is a deformed and fibered $A_2$ given by
$$uv=(z+m(x+a))z(z-2m(x-a)) \eqno(10)$$
We wrap one D5 brane on each of the two nodes of the singularity which are at $x=a$ and $x=-a$. The world--volume theory is a $U(1)_1 \times U(1)_2$ gauge theory with two gauge
singlets $\phi_1,\phi_2$ and two charged fields $Q_{12},Q_{21}$ with charges $1,-1$ and $-1,1$ respectively. The superpotential is given by
$$W={m \over 2} (\phi_1+a)^2+ m(\phi_2-a)^2+ Q_{12}Q_{21}(\phi_2-\phi_1) \eqno(11)$$
The F--terms arising from $W$ are
$$F_{\phi_1}=m(\phi_1+a)-Q_{12}Q_{21} \eqno(12)$$
$$F_{\phi_2}=2m(\phi_2-a)+Q_{12}Q_{21} \eqno(13)$$
$$F_{Q_{12}}=(\phi_2-\phi_1)Q_{21} \eqno(14)$$
$$F_{Q_{12}}=(\phi_2-\phi_1)Q_{12} \eqno(15)$$
In addition there are two D-terms for the two Abelian gauge groups which impose the condition $|Q_{12}|=|Q_{21}|$ due to the opposite charges of these fields. (For simplicity we assume that
these VEVs are real in the following.)
We find that there are two isolated supersymmetric vacua given by (in addition to nonsupersymmetric metastable vacua common in such models[\SUS])
$$\phi_1=-a \qquad \phi_2=a \qquad Q_{12}=Q_{21}=0 \eqno(16)$$
and
$$\phi_1=\phi_2={a \over 3} \qquad Q_{12}Q_{21}={4 \over 3} ma \eqno(17)$$

Note that in the first vacuum given by eq. (16), the gauge group is $U(1)_1 \times U(1)_2$ and all matter fields are massive. Since the Abelian groups remain unbroken, topological excitations 
of this vacuum are free magnetic monopoles for each $U(1)$ with masses $m_i=8 \pi a/g_i^2$ with $i=1,2$. This vacuum corresponds to a point on the ``Coulomb branch'' of the theory 
(in ${\cal N}=2$ terminology) On the other hand, in the second vacuum given by eq. (17), these Abelian groups (actually $U(2)$) are spontaneously broken but $Q_{12}$ and $Q_{21}$ are massless. 
As a result, topological excitations of this vacuum are vortices with tension $T_s=8 \pi m a/3$. This vacuum corresponds to a point on the ``Higgs branch'' of the theory.
We see that the two vacua are quite different from each other not only at the perturbative but also at the nonperturbative level.

The existence of these two isolated vacua means that there is a domain wall which interpolates between them. The domain wall solution is given by 
$$Q_{12}=Q_{21}={1 \over A}\sqrt{{{4ma} \over 3}} e^{-2az}   \eqno(18)$$
where $A^2=e^{-4az}+e^{4az}$ and 
$$\phi_1=-{a \over 3} (1+2tanhz) \qquad \phi_2={a \over 3} (2+tanhz) \eqno(19)$$
Above we again assumed that the wall is transverse to the $z$ direction and located at $z=0$. The tension of the domain wall is $T_w=8ma^2/3$. 
As before this wall can be seen as a D5 brane that is wrapped on the nodes and stretched between the two nodes (from $x=-a$ to 
$x=a$ on $C(x)$). It can be shown that the wrapping and stretching contribute factors of $4ma/3$ and $2a$ to the tension. Since this domain wall is not BPS, its solution and tension 
get corrections. Nevertheless, we expect it to be topologically stable due to the fact that it interpolates between two isolated vacua.

We found that the domain wall interpolates between two vacua with quite different physics; one that supports free monopoles and another that supports vortices. 
Two domain walls seperated by the second vacuum can be connected by vortices. This again is reminiscent of string theory in which D--branes are connected by 
fundamental strings. In this field theory, domain walls and vortices are similar to D--branes and fundamental strings repectively. Unfortunately,
since only one of the two vacua supports vortices, the walls are connected by vortices
on one side and not the other. Thus, these domain walls are quite different from D--branes. If we want domain walls connected by vortices on both sides, we need a theory in which
there are at least two different vacua that support vortices, which is necessarily a theory with more than two vacua.

It is easy to see that higher singularities, $A_n$ with $n>2$, lead to a larger number of isolated vacua and therefore to a larger number of domain walls with different properties. 
Consider the deformed $A_3$ singularity fibered over $C(x)$ and given by
$$uv=(z+2m(x-a))z(z+2m(x-a/2)(z+4m(x+a/4)) \eqno(20)$$
with three nodes at $x=a,a/2,-a$. We wrap one D5 brane on each one of the nodes and obtain a $U(1)_1 \times U(1)_2 \times U(1)_3$ gauge group with three singlets 
$\phi_1,\phi_2,\phi_3$ and the charged fields $Q_{12},Q_{21},Q_{23},Q_{32}$ with charges $(1,-1,0),(-1,1,0),(0,-1,1),(0,1,-1)$ under the three Abelian gauge groups respectively. 
In this case, the superpotential is
$$W=m(\phi_1-a)^2+m(\phi_2-a/2)^2+m(\phi_3+a)^2+Q_{12}Q_{21}(\phi_2-\phi_1)+Q_{23}Q_{32}(\phi_3-\phi_2) \eqno(21)$$
The F--terms obtained from the above superpotential are
$$F_{\phi_1}=2m(\phi_1-a)-Q_{12}Q_{21} \eqno(22)$$
$$F_{\phi_2}=2m(\phi_2-a/2)+Q_{12}Q_{21}-Q_{23}Q_{32} \eqno(23)$$
$$F_{\phi_3}=2m(\phi_2+a)+Q_{23}Q_{32} \eqno(24)$$
$$F_{Q_{12}}=Q_{21}(\phi_2-\phi_1) \eqno(25)$$
$$F_{Q_{21}}=Q_{12}(\phi_2-\phi_1) \eqno(26)$$
$$F_{Q_{23}}=Q_{32}(\phi_3-\phi_2) \eqno(27)$$
$$F_{Q_{32}}=Q_{23}(\phi_3-\phi_2) \eqno(28)$$
In addition, there are three D--terms arising from the three $U(1)$ gauge groups which vanish if $|Q_{12}|=|Q_{21}|$ and $|Q_{23}|=|Q_{32}|$. The above theory has four supersymmetric
vacua given by
$$(i) \quad \phi_1=\phi_2={3 \over 4}a \qquad \phi_3=-a \qquad Q_{12}=-Q_{21}=\sqrt{{{ma} \over 2}} \qquad Q_{23}=Q_{32}=0 \eqno(29)$$
$$(ii) \quad \phi_1=a \qquad \phi_2=\phi_3=-{a \over 4} \qquad Q_{12}=Q_{21}=0 \qquad Q_{23}=-Q_{32}=\sqrt{{{ma} \over 2}} \eqno(3)$$
$$(iii) \quad \phi_1=a \qquad \phi_2={a \over 2} \qquad \phi_3=-a \qquad Q_{12}=Q_{21}=Q_{23}=Q_{32}=0 \qquad \eqno(31)$$
$$(iv) \quad \phi_1=\phi_2=\phi_3={a \over 6} \qquad Q_{12}=-Q_{21}=\sqrt{{{5ma} \over 3}} \qquad Q_{23}=-Q_{32}=\sqrt{{{7ma} \over 3}} \eqno(32)$$

In vacuum (i) the gauge group is $U(1)_3$ with a spontaneously broken $U(2)$ and massless $Q_{12}$ and $Q_{21}$. The topological excitations are monopoles of $U(1)_3$ and two types of
non--Abelian vortices (with the same tension) due to the broken $U(2)$. Vacuum (ii) is very similar with the gauge group $U(1)_1$, spontaneously broken $U(2)$ and massless $Q_{23},Q_{32}$. 
Monopoles and non--Abelian vortices 
are again the topological excitations under the unbroken and broken gauge groups. In vacuum (iii), the gauge group is $U(1)_1 \times U(1)_2 \times U(1)_3$, all matter fields are massive 
and the topological excitations are
monopoles of each $U(1)$ factor. Finally in vacuum (iv) there is a spontaneously broken $U(3)$, with all $Q_{ij}$ massless and two types of non--Abelian vortices with different tensions. 
We see that these four vacua have 
very different perturbative and nonperturbative properties. In ${\cal N}=2$ terminology, vacua (i) and (ii) correspond to points in the ``mixed Higgs--Coulomb branches'' whereas vacua 
(iii) and(iv) correspond to points in the ``Coulomb'' and ``Higgs branches'' respectively.

There are six domain walls that interpolate between any two of these four vacua. For any two of the vacua (i)-(iv), we can write down a domain wall solution which generalizes
eq. (20). For example, consider the domain wall that interpolates between vacua (i) and (iii) above. Its solution is given by (with $Q_{23}=Q_{32}=0$ and $\phi_3=-a$)
$$Q_{12}=-Q_{21}={1 \over A}\sqrt{{{ma} \over 2}} e^{-az/2} \qquad \phi_1={a \over 8} (7+tanhz)  \qquad \phi_2={a \over 8} (5-tanhz)  \eqno(33)$$
where $A^2=e^{-az}+e^{az}$ and we assumed as before that the wall is transverse to the $z$ direction and located at $z=0$. The tension of the domain wall is $T_w=ma^2/2$. 
Since this domain wall is not BPS, its solution and tension receive corrections. However, we expect that it is topologically stable since it connects two different vacua.
The other five domain wall solutions are similar and we will not describe them in detail here.

Note that vacua (i) and (ii) both support (non--Abelian) vortices as topological excitations. Therefore, there is a configuration of multiple domain walls interpolating between these two 
vacua that can support vortices. Now, as opposed to the case of the $A_2$ singularity, these walls can be connected by vortices on both sides which looks like a configuration of D--branes 
connected by strings. Note however, that unlike that of a pair of D--branes, this configuration is not BPS due to ${\cal N}=1$ supersymmetry.
The vortices in both vacua (both sides of the walls) have the same tension, $T=\pi ma$ but may or may not carry the same $U(1)$ flux. In the present case, the vortices are identical
if they are both charged under the $U(1)_2$ subgroup of both $U(2)$s. Otherwise, the two vortices on either side of the walls are different even though they have the same tension.
These vortices correspond to D3 branes wrapped on the second node and stretched along the world--volume[\LAST,\COS,\CST].

The discussion can be easily generalized to D5 branes on $A_n$ (with $n>3$) singularities. In this case, there is a vacuum with a spontaneously broken $U(n)$ gauge group that corresponds 
to a point in the ``Higgs branch''. This vacuum supports $n$ types of non--Abelian vortices. Another vacuum has the gauge group $U(1)^n$ which is a point in the ``Coulomb branch''. 
This vacuum supports $n$ types of monopoles. 
In addition, there are a large number of vacua that correpond to points in the ``mixed Higgs--Coulomb branch''. These have gauge groups $U(1)^m$ and spontaneously broken $U(k_i)$ 
(where $m+\sum k_i=n$).
These vacua support $m$ types of monopoles and $k_i$ types of non--Abelian vortices. In this case, a configuration of multiple domain walls can be connected by many different types of vortices.
We see that, for D5 branes on $A_n$, the physics of multiple domain walls becomes quite interesting with different types of strings and monopoles between them.

\bigskip
\centerline{\bf 4. Conclusions and Discussion}
\medskip

In this paper, we showed that domain walls can live on resolved or deformed $A_n$ singularities. We found that BPS domain walls live on singularities which are resolved but not deformed. 
On the other hand, non--BPS domain walls live on deformed singularities which are also fibered over $C(x)$. The walls interpolate between vacua that are different perturbatively and 
nonperturbatively. In particuar some vacua have Abelian
groups and support monopoles whereas others have spontaneously broken groups that support vortices. We found that in the deformed $A_2$ case, domain walls are connected by vortces only 
on one side. In the deformed $A_3$ case, due to the fact that there are two different vacua that support vortices, walls can be connected by vortices on both sides. These vortices,
though they have the same tension, may or may not be identical depending on the flux they carry.

It is interesting to note that, a configuration with domain walls that are connected by vortices is very similar to that of D--branes in string theory (which are also connected by
fundamental strings). This similarity, in theories with ${\cal N}=2$ supersymmetry, has been noticed previously in refs. [\SHI-\ASY]. In such a framework, domain walls and vortices 
are analogous to D--branes and fundamental strings. It seems that
these field theories, in certain limits, have properties that are similar to string theory. It has been shown that, vortices are excitations of the domain
walls and their ends are charged under the wall gauge theory. In order to push this analogy further, one needs to show other similarities with string theory such as
the existence of other brane--like solitons in field theory, e.g. monopoles analogous to zero branes, symmetry enhancement when two domain 
walls overlap etc. Clearly this ``string theory'' limit of field theory in question corresponds to noncritical string theory since the world--volume theory does not contain gravity. 
Our results indicate that the above similarity between field and string theories 
can be extrapolated to theories with ${\cal N}=1$ supersymmetry. Even though such configurations with domain walls and vortices are no longer BPS, topological considerations imply
that they exist. 

A natural extension of our results is a description of the world--volume theories of the domain walls described above. This has been done for domain 
walls that arise in intersecting brane configurations[\AMI-\METO]. For domain walls that live on singularities, we expect to find similar but somewhat different 
world--volume theories since our brane configurations are not exactly dual to those in refs.[\AMI-\METO]. The most important difference is the amount of supersymmetry: 
the theories in refs. [\AMI-\METO] have ${\cal N}=2$ whereas those in section 3 have ${\cal N}=1$ supersymmetry. As a result, there is less control over quantum properties of domain 
walls compared to those in theories with ${\cal N}=2$ supersymmetry. In addition, even the theories with ${\cal N}=2$ supersymmetry in section 2 are not exactly dual to those
in refs. [\AMI,\MET,\METO]. In those theories, D6 branes give rise to fundamentals whereas 
the duals of the above brane configurations when described as intersecting branes, include semi--infinite D4 branes rather than D6 branes. 

\bigskip

\centerline{\bf Acknowledgements}

I would like to thank the Stanford Institute for Theoretical Physics for hospitality.

\vfill

\refout

\end
\bye